\documentstyle[preprint,prc,aps,amssymb]{revtex}
\begin{document}

\tightenlines

\draft

\title{\bf Unusual statistics of interference effects in neutron
scattering from compound nuclei}

\author{J. C. Berengut, V. V. Flambaum and G. F. Gribakin\cite{affil}}
\address{School of Physics, University of New South Wales,
Sydney 2052, Australia}

\date{\today}

\maketitle

\begin{abstract}
We consider interference effects between $p$-wave resonance scattering
amplitude and background $s$-wave amplitude in low-energy neutron scattering
from a heavy nucleus which goes through the compound nucleus stage. The first
effect is in the difference between the forward and backward scattering
cross sections (the ${\bf p}_i \cdot {\bf p}_f$ correlation). Because of the
chaotic nature of the compound states, this effect is a random variable
with zero mean. However, a statistical consideration shows that the
probability distribution of this effect does not obey the standard central
limit theorem. That is, the probability density for the effect averaged
over $n$ resonances does not become a Gaussian distribution with the
variance decreasing as $n^{-1/2}$ (``violation'' of the theorem!). We derive
the probability distribution of the effect and the limit distribution
of the average. It is found that the width of this distribution
does not decrease with the increase of $n$, i.e., fluctuations are
not suppressed by averaging. Furthermore, we consider the $\bbox{\sigma}
\cdot ({\bf p}_i \times {\bf p}_f )$ correlation and find that this effect,
although much smaller, shows fluctuations which actually increase
upon averaging over many measurements. This behaviour holds for
$\epsilon > \Gamma _p$ where $\epsilon $ is the distance to the resonance,
and $\Gamma _p$ is the resonance width. Limits of the effects due to finite
resonance widths are also considered. In the appendix we present a simple
derivation of the limit theorem for the average of random variables with
infinite variances.
\end{abstract}

\vspace{1cm}

%*******************************************************************

\section{Introduction}\label{intro}

Due to chaotic nature of compound nuclei, positions of $s$ and $p$
resonances in neutron scattering from a heavy nucleus, and amplitudes
involving these states, are uncorrelated. This gives rise to an unusual
statistical effect in the asymmetry of the transmission of neutrons
with positive and negative helicities \cite{Flambaum94}. This asymmetry
corresponds to the $\bbox{\sigma} \cdot {\bf p}$ correlation.
It violates parity conservation, and is produced by the weak interaction
in the nucleus, which mixes the $s$ and $p$ neutron partial waves.
The magnitude of the asymmetry is strongly enhanced if the neutron energy
is tuned into the $p$ resonance \cite{Sushkov}. In this case its magnitude
is determined by the perturbative mixing
\begin{equation}\label{mix}
\eta =\frac{\langle s|W|p\rangle }{E_s-E_p}
\end{equation}
of the $s$ and $p$ resonances by the weak interaction $W$. The matrix
element between the compound states behaves as a Gaussian random
variable, and $\eta $ is also a random variable with zero mean. The
characteristic mixing (and asymmetry) can be estimated simply as
$\eta _c \sim w/D$, where
$w$ is the root-mean-square matrix element, and $D$ is the mean spacing
between the compound nucleus resonances. However, when one takes a closer
look at the probability distribution of $\eta $, it turns out that its
variance is infinite! This effect originates from a high probability
to find small spacings $E_s-E_p$, which results in the slow decrease
of the probability density, $f(\eta )\propto 1/\eta ^2$.

The standard Central Limit Theorem (CLT) is not applicable to such random
variables. Instead, it can be shown that if we consider the statistics
of the average of $n$ such variables, the probability density
of the sum $(\eta _1+\dots +\eta _n)/n$ becomes independent of $n$ at large
$n$, i.e., fluctuations of $\eta $ are not averaged out \cite{limit}. This
contrasts the usual situation where CLT would give a Gaussian distribution
whose width decreases as $1/\sqrt{n}$. This unusual behaviour is explained by
the fact that among $n$ uncorrelated $\eta _i$ there is a large probability
to find one, whose magnitude is $n$ times greater than their typical value,
$\eta _i\sim n\eta _c$. Such $\eta _i$ will always dominate the sum
and ensure that fluctuations are not suppressed by averaging.
Another physical instance in which rare events dominate the distribution is
seen in ``Levy flights'' in the random force diffusion model \cite{levy}. In
a usual homogeneous system, diffusion is modelled by Brownian motion, where
the distribution of displacements is Gaussian. But in the random force
diffusion model, disorder induces rare but large displacements which dominate
the distribution (the Levy distribution). Under certain values of parameters,
these have statistics similar to those found in nuclear scattering problems.

In this paper we analyse other effects in neutron scattering that have
such unusual statistics. They are more conventional than those discussed
in Ref. \cite{Flambaum94}, since they do not involve the weak interaction.
For example, in Sec.~\ref{spinless} we consider the difference between the
differential cross-section in the forward and
backward scattering, due to the interference of the $p$-wave resonance
scattering and the background scattering amplitude,  for a spinless
particle. This effect can be described as the ${\bf p}_i\cdot {\bf p}_f$
correlation, where ${\bf p}_i$ and ${\bf p}_f$ are the momentum of the
incident and emitted particles, respectively. We derive the statistics of
the observable and the way it behaves upon averaging over many different
nuclei, in a similar fashion to what was done in \cite{Flambaum94}. We
also discuss the limit of this effect when finite widths of the resonances
are taken into account. In Section~\ref{spin} we rederive the
${\bf p}_i\cdot {\bf p}_f$ correlation for particles with spin (e.g., neutrons)
incident upon a spinless nucleus, as well as study the effect of a different
correlation between the neutron spin $\bbox{\sigma }$ and the scattering
plane, $\bbox{\sigma }\cdot ({\bf p}_i\times {\bf p}_f)$. This
is shown to have different statistics to the first correlation, and we derive
a limit theorem for the average of the second correlation effect (details
can be found in the Appendix). As it turns out, fluctuations of this average
increase upon averaging, because when larger sets of data are considered
there is a finite probability of finding an effect whose magnitude is
$\sim n^2$ times larger than its typical value.

%*******************************************************************

\section{Cross section asymmetry for a spinless particle}\label{spinless}

Let us first study the simple case of the ${\bf p}_i\cdot {\bf p}_f$
correlation for scattering of a spinless particle. Here we consider the
difference between forward and backward elastic scattering differential
cross sections near threshold,
due to interference of the $p$-wave resonance scattering with the
background $s$-wave amplitude. For a spinless particle the scattering
amplitude at low momenta is written using the Breit-Wigner formula as
(see, e.g., \cite{Landau:77})
\begin{equation}\label{eq-breit}
f( \theta ) =-A-\frac{g_{p}}{2k}\frac{\Gamma _{p}^{(n)}}
{E-E_p +\frac{i}{2}\Gamma _p}\cos \theta , 
\end{equation}
where $A$ is the $s$-wave scattering length, $k$ and $E$ are the wave number
and energy of the projectile, and $g_p$, $E_p$, $\Gamma _p^{(n)}$ and
$\Gamma _p$, are the statistical weight, energy, capture (or elastic) width
and total width of the $p$ resonance, respectively.
We assume that at energy $E$ the $s$-wave background is nonresonant, and there
is a $p$-wave resonance nearby, a condition which would favour larger
asymmetries. This leads to expressions for the forward ($+$) and backward
($-$) scattering amplitudes
\begin{equation}\label{eq-amp_spinless}
f^\pm =-A\mp \frac{g_{p}}{2k}\frac{\Gamma _{p}^{\left(
n\right) }}{\varepsilon }
\end{equation}
where $\varepsilon =E-E_{p}$ is the distance to the $p$-wave
resonance, and we assume that it is greater then the resonance width,
i.e., $\varepsilon \gg \Gamma _{p}$.

The relevant observable is the asymmetry
\begin{equation}\label{eq-observable}
x= \frac{(d\sigma /d\Omega )_+ -(d\sigma /d\Omega )_-}{(d\sigma /d\Omega )_+
+(d\sigma /d\Omega )_- },
\end{equation}
where $(d\sigma /d\Omega )_\pm$ are the forward and backward scattering
cross sections. Substituting amplitudes (\ref{eq-amp_spinless}) and taking
into account that at low momenta the contribution of the $p$ wave is much
smaller than that of the $s$ wave, we obtain
\begin{equation}\label{eq-corr1}
x=\frac{\Gamma _{p}^{(n)}}{\beta \varepsilon }
\end{equation}
where $\beta =Ak/g_p$. A typical value of this asymmetry is
$g_p \Gamma _{p}^{(n)}/AkD$.

%*******************************************************************

\subsection{Statistical analysis}\label{stat1}

We would now like to obtain the probability density for the observable $x$.
The capture width is proportional to the square of the capture amplitude.
The capture amplitudes for complex compound nuclear states have a
Gaussian distribution \cite{BM}, hence, the widths $\Gamma _{p}^{(n)}$ are
distributed according to the Porter-Thomas law
\begin{equation}\label{eq-porter}
g(\gamma ) =\frac{1}{\sqrt{2\pi \bar\gamma \gamma }}\exp \left(
-\frac{\gamma }{2\bar \gamma }\right) ~,
\end{equation}
where $\gamma \equiv \Gamma _{p}^{\left( n\right) }$ for convenience and
$\bar \gamma $ is the mean width $\bar\gamma =\int \gamma g(\gamma )d\gamma $.

For a given energy $E$ the distance to the nearest $p$ resonance in a compound
nucleus is random. If the relative positions of the $p$ resonances were
uncorrelated it would be described by a Poissonian distribution
\begin{equation}\label{eq-poisson}
f_D(\varepsilon )=D^{-1}\exp \left( -\frac{2\left|
\varepsilon \right| }{D}\right)
\end{equation}
where $D$ is the mean spacing between the $p$ resonances and
$\overline{|\varepsilon |} =D/2$. Correlations between the positions of
compound states of the same symmetry, often referred to as
level repulsion, modify the above distribution. These correlations
are described by the random matrix theory \cite{Brody}, and can be
approximated by the Wigner law. In this case the distance to the nearest 
$p$ resonance has the following probability density \cite{Flambaum94}:
\begin{equation}\label{eq:wigner}
f_D(\varepsilon )=D^{-1}\exp \left( -\frac{\pi \varepsilon ^2}{D^2}\right)~.
\end{equation}
To avoid confusion we should stress that here $\varepsilon $ is the distance
to the nearest $p$-wave resonance and not the interval between the $p$-wave
resonances. Therefore, Eq. (\ref{eq:wigner}) differs from the Wigner-Dyson
distribution. The difference between Eqs. (\ref{eq-poisson}) and
(\ref{eq:wigner}) is not very important for our consideration, as long as
$f_D(\varepsilon )$ remains finite (and equal to $D^{-1}$) at
$\varepsilon \rightarrow 0$ \cite{Flambaum94}.

Using Eqs. (\ref{eq-porter}) and (\ref{eq:wigner}) we calculate the
distribution of the observable $x$ as
\begin{eqnarray}
f(x) &=&\int_0^\infty d\gamma \int_{-\infty }^{\infty} d\varepsilon
f_D( \varepsilon ) g(\gamma ) \delta \left( x-\frac{\gamma }
{\beta \varepsilon }\right) \nonumber \\
&=&\frac{1}{\sqrt{2\pi x_0 |x|}}\int _0^\infty \sqrt{t} \exp \left(
-\frac{|x|t}{2x_0}-\pi t^2 \right) dt ,
\label{eq-fdist1}
\end{eqnarray}
where $x_0=|\bar \gamma /\beta D|$ represents some typical value of the effect.
The integral in (\ref{eq-fdist1}) can be given explicitly in terms of the
parabolic cylinder functions $D_\nu (z)$,
\begin{equation}\label{eq:fdist11}
f(x)=\frac{(2\pi )^{-3/4}\Gamma (\frac{3}{2})}{\sqrt{2\pi x_0 |x|}}
\exp \left( \frac{x^2}{32\pi x_0}\right) D_{-\frac{3}{2}}
\left( \frac{|x|}{x_0\sqrt{8\pi }}\right)~.
\end{equation}
On the other hand, one can easily find the asymptotic behaviour of the
probability density at $|x|\gg x_0$ directly from Eq. (\ref{eq-fdist1}):
\begin{equation}\label{eq-asy1}
f(x) \simeq x_0/x^2 .
\end{equation}
The CLT will not apply to distributions with this asymptotic form,
since they do not have a finite variance: $\int x^2f(x)dx\rightarrow \infty $.

%*******************************************************************

\subsection{Limit theorem for the first correlation}\label{limit1}

Suppose the forward-backward asymmetry is measured in an experiment
where a number of different nuclei with similar-sized cross sections
are involved. The measurement will then yield some average asymmetry,
and we want to find the probability distribution of it. Otherwise,
one may just analyse the asymmetries measured separately for a number
of nuclei. Let us then consider the average of $n$ independent random
variables $x_i$ introduced above,
\begin{equation}\label{eq:X}
X=\frac{1}{n}\sum _{x=1}^{n}x_i .
\end{equation}
In Ref. \cite{Flambaum94} a derivation of the limit theorem for
distributions with asymptotic behaviour (\ref{eq-asy1}) was presented.
A general solution of the problem of limit distributions of sums of
independent variables with an infinite variance for which
$f(x)\propto |x|^{-\alpha -1}$, can be found in Ref. \cite{limit} ($\alpha >0$
to keep the total probability $\int f(x)dx$ finite). A simple derivation
of the limit theorem for such distributions is given in the Appendix.

The random variable $x$ has a symmetric probability distribution, $f(-x)=f(x)$.
In this case for $\alpha =1$ the limit distribution is obtained from
Eqs. (\ref{eq:Fnfab}) and (\ref{eq:Cauchy}) with $a=0$ and $c=\pi x_0$. So, for
$n\rightarrow \infty $ the probability density $F_n(X)$ approaches its limit
form
\begin{equation}
F_n(X)=\frac{1}{\pi}\frac{X_c}{X^2+X_c^2},
\end{equation}
where $X_c=\pi x_0$ \cite{Flambaum94}. This is called the Cauchy distribution,
and its main property is that it is independent of $n$, in particular, it
does not become narrower as $n$ increases. Therefore, fluctuations are not
suppressed by averaging. Compare this with the standard central limit theorem,
where the width decreases as $\sigma_n = \sigma_{1}/\sqrt{n}$.

The parameter $X_c$ for our physical observable is given by
$X_c=\pi \bar \gamma /\beta D=\pi g_p\overline{\Gamma _p^{(n)}}/AkD$.
Throughout
the derivation we considered the scattering length $A$ as a constant.
Indeed, if the energy $E$ does not coincide with an $s$-wave resonance, the
$s$-wave amplitude $-A$ represents the potential scattering amplitude. It
does not vary strongly between isotopes, or nuclei of similar masses, because
the nuclear potential does not vary much. The energy scale of its variation
is MeV, similar to single-particle energy level spacing. Contrast this with
the  scale of the compound resonance spacings which are of the order of
10 eV. This difference in energy scales means that the compound resonance
states can be treated statistically, while the scattering length is treated
as a constant.

%*******************************************************************

\subsection{Influence of the resonance widths on the statistics
of the average asymmetry}
\label{widths}

The above calculations have been based on the assumption that
$\varepsilon \gg \Gamma _{p}$, so that the possible effects of the resonance
widths have been neglected. This is justified, as long as the probability
of finding a very small interval $\varepsilon \sim \Gamma _p$ is indeed
small. However, it is easy to see the role of the width as we increase
$n$. As we explained in the Introduction, the nonvanishing fluctuations
of the average depends on having one value of the effect large,
$x_{i}\sim nx_{c}$, where $x_{c}$ is a typical value. Indeed, we can expect
that if we make $n$ measurements then at least one will have an energy
spacing of the order $\varepsilon \sim D/n$, thus giving a large
asymmetry (\ref{eq-corr1}). The energy denominator, however, can not be made
arbitrarily small, and as smaller $E-E_p$ are considered, it will reach a
limit $| \varepsilon +i\Gamma _{p}/2| \sim \Gamma _{p}$. Thus
$D/n\sim \Gamma _{p}$ determines the largest values of $n$,
beyond which the Cauchy distribution of the average effect begins to turn
into a Gaussian one. Hence our statistical analysis is valid until
$n\sim D/\Gamma _{p}$ (in heavy non-fissionable nuclei
$D/\Gamma _{p}\sim 500$).

If we continue to take measurements after this and further increase
the number of measurements $n$, the maximum value of the asymmetries will
stabilise, being of the order $x_{i}\sim x_0 D/\Gamma _{p}=
\pi \bar\gamma /\beta \Gamma _{p}$. Hence we will no longer have increasingly
large values of the effect to continue the ``non-vanishing'' averaging. Thus,
when $n\gg D/\Gamma _{p}$ the Gaussian statistics take over, the standard
central limit theorem applies, and the usual $1/\sqrt{n}$ suppression of
fluctuations takes place. Note that $x_0 D/\Gamma _{p}\gg x_0$ in fact
determines the true finite, but large, variance of $x$. Beyond this value
$f(x)$ decreases faster than $x^{-2}$, and it effectively determines
the lower and upper limits in the variance integral $\int x^2f(x)dx$.

%*******************************************************************

\section{Spin one-half particle correlations}\label{spin}

Let us now consider scattering of low-energy spin-$\frac{1}{2}$ particles,
looking at both the ${\bf p}_i\cdot {\bf p}_f$ and $\bbox{ \sigma }
\cdot ({\bf p}_i\times {\bf p}_f)$ correlations. Again we assume that
there is no nearby compound nucleus $s$-wave resonance and that the asymmetry
is dominated by one nearby $p$-wave resonance. This is justified because
the statistics of the average effect $X$ for large $n$, $F_n(X)$, is
determined by large values of the individual effects, i.e., by the asymptotic
large-$x$ behaviour of the probability density $f(x)$, cf. Eq. (\ref{eq-asy1})
\cite{comment}.

Consider scattering of the neutron, $s=\frac{1}{2}$ from a nucleus with spin
$I$. The total angular momentum of the $p$-wave neutron is ${\bf j}={\bf l}+
{\bf s}$ and the total angular momentum of the compound resonance is
${\bf J}={\bf I}+{\bf j}$.  The amplitude of $p$-wave resonant
scattering at arbitrary angle $\theta $ between the incoming and outgoing
particle can be written as (see Ref. \cite{Flambaum83})
\begin{eqnarray}
f_{p} &=&-\frac{1}{2k}\sum_{{j,j_{z},m}\atop {j', j_z',m'}} C_{II_{z}j'j_z'}
^{JJ_{z}}C_{1m'\frac{1}{2}s_z'}^{j'j_z'} \sqrt{4\pi }Y_{1m'}^{*}({\bf n}_k')
\sqrt{\Gamma _{p_{j'}}^{(n)}(E)}\nonumber \\
&&\times \frac{1}{E-E_p+\frac{i}{2}\Gamma _p}C_{II_zjj_z}^{JJ_z}
C_{1m\frac{1}{2}s_z}^{jj_z}\sqrt{4\pi }Y_{1m}({\bf n}_{k})\sqrt{\Gamma
_{p_j}^{(n)}(E)}, \label{eq-modBreit}
\end{eqnarray}
where $I_z$, ${\bf n}_k$, $s_z$, and  $I_z'$, ${\bf n}_k'$, $s_z'$,
describe the projection of the target spin, the direction of the
neutron momentum and the projection of the neutron spin in the initial and
final states, respectively, $\Gamma _{p_{j}}^{(n)}$ is the capture width for
the neutron with angular momentum $j$, the $Y_{lm}$ are the angular
wave functions, and $C_{II_zjj_z}^{JJ_z}$ are the Clebsch-Gordon coefficients.

Consider scattering of a neutron incident along the $x$ direction off a
spinless $\left( I=0\right) $ nucleus. We quantise the neutron spin in the
$z$ direction and consider neutrons scattered in the $xy$ plane.
The $s$-wave scattering amplitude is simply $-A\delta _{s_zs_z'}$. Having
in mind that we need to calculate interference terms between the $s$ and
$p$ waves, we can write the $p$-wave scattering amplitude (\ref{eq-modBreit})
in the following form
\begin{equation}\label{eq-amp_spin}
f_p(\theta ) =-\frac{1}{2k}\sum_{m}\left| C_{1m\frac{1}{2}
s_{z}}^{jj_{z}}\right| ^{2}Y_{1m}^{*}({\bf n}_k') \frac{1}
{E-E_{p}+\frac{i}{2}\Gamma _p}Y_{1m}({\bf n}_k) 4\pi \Gamma _{p_j}^{(n)}(E).
\end{equation}

%*******************************************************************

\subsection{First correlation}\label{corr1}

The resonance $p$-wave amplitude for forward and backward scattering
(see Sec. \ref{spinless}) with $j=\frac{1}{2}$ is
\begin{equation}\label{eq:p1/2}
f_{p_{1/2}}^\pm =\mp \frac{1}{2k}\frac{\Gamma _{p_{1/2}}
^{(n)}}{E-E_p +\frac{i}{2}\Gamma _p},
\end{equation}
which is similar to the spinless particle scattering (Eq. \ref
{eq-breit}), with the parameter $g_{p}=1$.
Similarly for $j=\frac{3}{2}$ states the amplitude is
\begin{equation}\label{eq:p3/2}
f_{p_{3/2}}^\pm =\mp \frac{2}{2k}\frac{\Gamma _{p_{3/2}}
^{(n)}}{E-E_p +\frac{i}{2}\Gamma _p}
\end{equation}
which is similar to spinless particle scattering with $g_p=2$.
This means that the statistics derived for the spinless
particle ${\bf p}_i\cdot {\bf p}_f$ correlation in Sec. \ref{spinless} are
valid for the case where spin is included. In fact, since we do not know
whether the nearest resonance is $p_{1/2}$ or $p_{3/2}$ we must combine the
two distributions.

%*******************************************************************

\subsection{Second correlation}\label{corr2}

The second correlation $\bbox{\sigma }\cdot ({\bf p}_i\times
{\bf p}_f)$ between the direction of the spin relative to the scattering plane
is, of course, specific to particles with a non-zero spin. To calculate the
asymmetry of the cross section with respect to flipping the spin, we take the
initial neutron momentum along the
$x$-direction as before, and look at the difference between the scattering
amplitude in the $+y$ direction, $f^+$, and that in the $-y$ direction,
$f^-$. Equation (\ref{eq-amp_spin}) yields the $p$-resonance scattering 
amplitudes in the $+y$ and $-y$ directions for $j=\frac{1}{2}$
\begin{equation}\label{eq:fp1/2}
f_{p_{1/2}}^\pm =\pm \frac{i}{2k}\frac{\Gamma _{p_{1/2}}^{(n)}}
{E-E_p+\frac{i}{2}\Gamma _p},
\end{equation}
and similarly for $j=\frac{3}{2}$ we obtain
\begin{equation}\label{eq:fp3/2}
f_{p_{3/2}}^\pm =\mp \frac{i}{2k}\frac{\Gamma _{p_{3/2}}^{(n)}}
{E-E_p +\frac{i}{2}\Gamma _p}
\end{equation}
which differs from the $j=\frac{1}{2}$ case only by sign.

Thus the total scattering amplitude for the $\bbox{\sigma }\cdot ({\bf p}_i
\times {\bf p}_f)$ correlation is given by
\begin{equation}\label{eq-totalamp2}
f^\pm =-A\mp \frac{i\eta_{p}}{2k}\frac{\Gamma _{p}^{\left(
n\right) }}{E-E_{p}+\frac{i}{2}\Gamma _{p}}  
\end{equation}
where $\eta_p=-1$ for $j=\frac{1}{2}$ and $\eta_p=+1$ for $j=\frac{3}{2}$.
Then,
taking into account that the second term in Eq. (\ref{eq-totalamp2}), which
represents the $p$-wave contribution, is much smaller than the first one, we
obtain for the observable difference of the corresponding cross sections
[see Eq. (\ref{eq-observable})]
\begin{equation}\label{eq:2nd}
x=\frac{\eta_p}{2kA} \frac{\Gamma _p \Gamma _p^{(n)}}
{(E-E_p)^2 +\frac{1}{4}\Gamma _p^2}.
\end{equation}
As we discussed above, the scattering length varies weakly. The same is
true for the total width of the compound resonances $\Gamma _p$. Its
fluctuations are small because it is dominated by the radiative width, given
by the a sum of a large number of partial widths due to transitions into all
lower-lying nuclear states. Introducing $\beta =2kA/\eta_p\Gamma _p$, and
taking into account that $\varepsilon =E-E_p$ is usually much larger than the
resonance width, $|\varepsilon | \gg \Gamma _{p}$, we obtain for the asymmetry
\begin{equation}\label{eq-corr2}
x=\frac{\Gamma _{p}^{(n)}}{\beta \varepsilon ^2}.
\end{equation}
The typical size of this effect $\eta_p\Gamma _{p}^{(n)}\Gamma _p/AkD^2=
(\eta_p\Gamma _{p}^{(n)}/AkD)(\Gamma _p/D)$, is much smaller than the
first correlation, by a factor of $\Gamma _p/D$. However, this observable has
a $\varepsilon ^{-2}$ dependence on the distance to the nearest $p$
resonance, while the first correlation was proportional to $\varepsilon ^{-1}$.
The $\varepsilon ^{-2}$ singularity emphasizes even stronger the possibility
of small denominators. Note also, that for a given scattering length $A$ the
sign of this interference effect is always the same. We will see that this
leads to a very different statistics of the
$\bbox{\sigma }\cdot ({\bf p}_i\times {\bf p}_f)$ correlation.

%*******************************************************************

\subsubsection{Statistical analysis}\label{stat2}

Let us derive a probability distribution for the observable $x$ given by
Eq. (\ref{eq-corr2}). Similarly to Sec. \ref{stat1} we have
\begin{equation}\label{eq-fcalc2}
f(x) =\frac{1}{\sqrt{2\pi \bar \gamma }}\int_0 ^\infty
\int _{-\infty }^\infty \exp \left( -\frac{\pi \varepsilon ^2}
{D^2} -\frac{\gamma }{2\bar \gamma }
\right) \frac{d\varepsilon d\gamma }{\sqrt{\gamma }}
\delta \left( x-\frac{\gamma }{\beta \varepsilon ^2}\right)
\end{equation}
where we again use $\gamma $ for the capture width $\Gamma _{p}^{(n)}$, and
the probability densities $g(\gamma )$ and $f_{D}( \varepsilon )$ are
taken from Eqs. (\ref{eq-porter}) and (\ref{eq:wigner}), respectively.
Assuming that the scattering length is positive, hence, $\beta >0$, we
calculate the above integral and obtain
\begin{equation}\label{eq-fcalc21}
f(x) =\frac{\sqrt{x_0}}{\sqrt{\pi x}(x+\pi x_0)}~,\quad x>0,
\end{equation}
and $f(x)=0$ for $x<0$, where
\begin{equation}\label{eq:corr2x0}
x_0=\frac{2\bar \gamma }{\beta D^2}
\end{equation}
characterises typical values of the asymmetry (\ref{eq-corr2}).
The asymptotic behaviour of this probability density at $x\gg x_0$ is
\begin{equation}\label{eq:asy2}
f(x) \simeq \frac{\sqrt{x_0}}{\sqrt{\pi }|x|^{3/2}} ~.
\end{equation}
The probability density $f(x)$ is normalized as $\int _0^\infty f(x)dx =1$.
However, the corresponding mean value $\int f(x)x dx$ is infinite, and
the integral for the variance $\int f(x) x^2 dx$ diverges even faster than
that for the first correlation [cf. Eq. (\ref{eq-asy1})]. This signifies
even larger fluctuations of the second correlation effect.

%*******************************************************************

\subsubsection{Limit theorem for the second correlation}\label{limit2}

The probability distribution of the second correlation (\ref{eq:asy2})
corresponds to $\alpha =\frac{1}{2}$ (see Appendix). Using the asymptotic
parameters $c_1=0$ and $c_2=\sqrt{x_0/\pi }$ [compare Eqs. (\ref{eq:asy2}) and
(\ref{eq:asymasym})] we obtain $c=\sqrt{2x_0}$ and $\beta =1$ from Eqs.
(\ref{eq:ccc}) and (\ref{eq:beta}). In fact, it is possible to calculate
the Fourier transform of $f(x)$ of Eq. (\ref{eq-fcalc21}) explicitly,
\begin{eqnarray}\label{eq:fom1/2}
\tilde f(\omega )&=&e^{i\pi x_0\omega }\pi ^{-1/2}\Gamma ( 1/2,
i\pi x_0 \omega ) \nonumber \\
&=& e^{i\pi x_0\omega }\left[ 1-\sqrt{2x_0}(1\pm i)|\omega |^{1/2}
+O(\omega ^{3/2})\right] ~,
\end{eqnarray}
where $\Gamma (\dots )$ is the incomplete $\Gamma $-function, and $\pm $
corresponds to $\omega \gtrless 0$.

It follows now from Eqs. (\ref{eq:Fnfab}) and (\ref{eq:smirnov}) that the
limit distribution $F_n(X)$ of the average effect $X$ is given by
\begin{equation}\label{eq:second}
F_n(X)=\sqrt{\frac{nx_0}{\pi }}\,\frac{e^{-nx_0/X}}{X^{3/2}}~,\quad (X>0).
\end{equation}
This equation shows explicitly that as the number of effects included in the
average increases, the distribution widens proportionally to $n$. Accordingly,
the typical values of the average also grow as $X\sim nx_0$.

To understand this recall that the second asymmetry is inversely proportional
to $\varepsilon ^2$. Thus, when we consider the average of $n$ such variables,
one of them is likely to have $\epsilon \sim D/n$, which makes it $n^2$
times greater than the typical value $x_0$. This variable will dominate
the average and give $X\sim nx_0$. Also, to describe a possible experiment
more realistically, one must combine the distributions with different
$\eta_p$ remembering that the probability to find a close $p_{3/2}$ resonance
is twice that of a $p_{1/2}$ resonance.

The role of resonance widths can be understood in the same way as it was done
in Sec.~\ref{widths}. The limit statistics (\ref{eq:second}) is valid until
$n \sim D/\Gamma_p$, i.e., when the characteristic value of the
maximal effect that dominates the average $X$ requires denominators
as small as $\varepsilon \sim D/n \sim \Gamma $. When
$n \gg D/\Gamma_p$ the usual statistics of the Central Limit
Theorem become valid, and we eventually have typical values of the average
decreasing as $1/\sqrt{n}$.

%*******************************************************************

\section{Conclusion}\label{conc}

We have considered interference effects between $p$-wave resonance neutron
scattering amplitude and background $s$-wave amplitude in compound nuclei and
found that these effects do not obey the Standard Central Limit Theorem. That
is, the probability density of the average effect over $n$ measurements,
$X=\frac{1}{n}\sum_{i=1}^{n}x_{i}$ , does not tend to a Gaussian distribution
with variance $\sigma_{n}^{2}=\sigma_1 ^2/n$ for large $n$. We have
examined two effects, (i) the ${\bf p}_i\cdot {\bf p}_f$ correlation, which
corresponds to the forward-backward asymmetry of the differential cross
section, and (ii) the $\bbox{\sigma }\cdot ({\bf p}_i\times {\bf p}_f)$
correlation, which describes the asymmetry of the scattering cross section
with respect to flipping the spin relative to the scattering plane.

The first of these was found to have a distribution with asymptotic
behaviour $f(x) \propto 1/x^2$ for large $x$. In this case the limit theorem
for the distribution of the average $X$ tends to a Cauchy distribution
$F_{n}(X) =X_c/[\pi (X^2+X_c^2)]$. This is independent of $n$, so the typical
value of the average ($X_{c}$) does not decrease with increasing number of
measurements. Physically this is understood by the following arguments. The
asymmetry in question is inversely proportional to the spacing between the
incident neutron energy and the energy of the closest $p$-wave resonance
($x\propto \varepsilon ^{-1}$). If we have $n$ measurements, we have a
high probability that one of these spacings will be of the order
$\varepsilon \sim D/n$ where $D$ is the mean $p$ level
spacing. This will produce an asymmetry of the size $x\sim nx_0$ where $x_0$
is a typical value of the effect. Thus the typical average value is $X\sim
x_0$, nonvanishing with increasing $n$.

The second correlation we considered produces a much smaller effect than the
first correlation. However, it has been found to have a $\varepsilon ^{-2}$
dependence, giving a distribution with asymptotic behaviour
$f(x) \propto x^{-\frac{3}{2}}$ for large $x$. This means that there is
a higher probability to obtain relatively large values of $x$, compared to
the first correlation. As a result, typical values for the average of $n$
asymmetries actually increase with increasing number of measurements $n$
as $X\sim nx_0$.

Above we assumed $\epsilon > \Gamma_p$ where $\epsilon $ is the distance to
the resonance and $\Gamma_p$ is the resonance width. When we consider the
the influence of the resonance widths, it is found that they affect the
distribution by limiting the size of the effect, $x$. Indeed, the minimum
value for the denominator $| \varepsilon +i\Gamma_p/2|$ is given by
$\sim \Gamma _p$, hence, the maximal possible effects are limited. We
have found that our analysis of the statistics of the averages $X$ is valid
for $n< D/\Gamma _p$ (for heavy non-fissionable nuclei
$D/\Gamma_p \sim 500$). For $n\gg D/\Gamma _p$ we expect the average to once
again obey the standard CLT and vanish with increasing $n$.

Because of the interest in scattering problems, it would be of benefit to
actually perform the experiments discussed in this paper. If one measured the
first correlation in low-energy neutron scattering off different isotopes
of heavy nuclei, then we expect to see an observable effect in the average.
This would not decrease until the number of measurements $n \gg 500$.
(Note, however, that the total number of relatively stable isotopes
$\sim 1000$). Because it is hard to measure the forward and backward
scattering, the experiment would have to be performed at some angle
$\theta$ to the axis of incidence. This would only cause an extra factor of
$\cos \theta $ in the value of the effect.

The second correlation is much smaller (by a factor $\Gamma_p/D \sim 10^{-3}$).
This would make it much harder to observe, but it would have an
increasing typical value upon averaging over many measurements. This means
that one might be able to observe the effect after performing the experiment
over many isotopes. 

%*****************************************************************************

\appendix

\section{Limit theorem for probability distributions with
infinite variances}\label{LimTh}

Consider a random variable whose probability density has the following
asymptotic behaviour:
\begin{eqnarray}\label{eq:asymasym}
f(x)=\cases {c_1/|x|^{\alpha +1},~x\rightarrow -\infty , \cr
c_2/x^{\alpha +1},~x\rightarrow +\infty , \cr }
\end{eqnarray}
with $0<\alpha <2$, and is normalized in the usual way, $\int f(x)dx=1$.
The existence of the mean, $\int xf(x)dx$, depends on whether $\alpha $ is
greater or less than unity, but the variance integral $\int x^2f(x)dx$ is
infinite in both cases, and the standard Central Limit Theorem is inapplicable.

To derive the limit statistics of the average $X=\frac{1}{n}\sum _{x=1}^{n}x_i$
of $n$ independent random variables $x_i$, we use characteristic functions
(or Fourier transforms)
\begin{equation}\label{eq:Fourier}
\tilde{f}(\omega )=\int _{-\infty }^\infty e^{-i\omega x} f(x)dx .
\end{equation}
The Fourier transform of the probability density $F_n(X)$ of the average
$X$ is given by
\begin{eqnarray}
\tilde F_n(\omega ) &=&\int _{-\infty}^{\infty }e^{-i\omega X}dX \int
\delta \left( X - \frac{1}{n}\sum _{i=1}^{n}x_i \right) \prod _{i=1}^{n}
f(x_i) dx_i \nonumber \\
&=& \prod _{i=1}^{n} \tilde f ( \omega /n)=
\left[\tilde f ( \omega /n ) \right]^n . \label{eq:Fnom}
\end{eqnarray}
Thus, the form of $F_n(\omega )$ for large $n$ is related to that
of $\tilde{f}(\omega )$ at small $\omega $. This is in turn decided by the
large-$x$ asymptotic behaviour of $f(x)$ given by Eq. (\ref{eq:asymasym}).

For $1<\alpha <2$ the integral $\int xf(x)dx$ converges and $\tilde f(\omega )$
can be written as
\begin{equation}\label{eq:expand}
\tilde f(\omega )=1-i\omega \int xf(x)dx +\int (e^{-i\omega x} -1+i\omega x)
f(x)dx .
\end{equation}
Let us consider the contribution of the interval from $0$ to $+\infty $ to
the last term above. Using the asymptotic form (\ref{eq:asymasym}) we
present it as
\begin{equation}\label{eq:interm}
\int _0^{\infty}(e^{-i\omega x} -1+i\omega x)\left[ f(x)-\frac{c_2}
{x^{\alpha +1}}\right]dx +c_2\int _0^{\infty}(e^{-i\omega x} -1+i\omega x)
\frac{dx}{x^{\alpha +1}}.
\end{equation}
If we assume that $f(x)$ approaches its asymptotic behaviour sufficiently
rapidly, e.g., $|f(x)-c_2/x^{\alpha +1}|\propto O(1/x^{\alpha +2})$, then
the first integral above behaves as $O(\omega ^2)$ at $\omega \rightarrow 0$.
To calculate the second integral we turn the integration path into the
complex plane by changing the variable $\omega x =it$ (for $\omega >0$ the
$t$ is real positive), which gives
\begin{equation}\label{eq:Gamint}
c_2e^{i\frac{\pi \alpha }{2}}
\omega ^\alpha \int _0 ^{\infty}t^{-\alpha -1}(e^{-t}-1+it)dt,
\end{equation}
where the integral is a representation of the $\Gamma $-function on a segment
of the negative argument axis,
$\Gamma (-\alpha )=-\Gamma (-\alpha +1)/\alpha $ \cite{comment1}.

Applying the same procedure to the integral over $(-\infty , 0)$ in expression
(\ref{eq:expand}), and turning the integration path into the complex plane by
using $\omega x =-it$ (for $\omega >0$ and real positive $t$), we obtain the
expansion of $\tilde f(\omega )$ at small $\omega $:
\begin{equation}\label{eq:semif}
\tilde f(\omega )=1-i\omega a-\left( c_1e^{-i\frac{\pi \alpha }{2}}
+c_2e^{i\frac{\pi \alpha }{2}}\right)\omega ^\alpha \frac{\Gamma (\alpha -1)}
{\alpha }+\dots 
\end{equation}
where $a=\int xf(x)dx$ is the mean value. The expansion for negative $\omega $
can be obtained from the above by simply replacing $\omega $ with $|\omega |$
and complex-conjugating the exponential phase factors in the second term.
Finally, at the same level of accuracy, we can re-write expansion
(\ref{eq:semif}) in the form valid for positive and negative small
$\omega $:
\begin{equation}\label{eq:final}
\tilde f(\omega )\simeq e^{-i\omega a}\left[ 1-c \left( 1+i \beta \,\mbox{sign}
(\omega ) \tan \frac{\pi \alpha }{2} \right) |\omega |^\alpha \right]~,
\end{equation}
where
\begin{eqnarray}
c&\equiv &(c_1+c_2)\frac{\Gamma (1-\alpha )}{\alpha } \cos
\frac{\pi \alpha }{2}~,\quad c >0, \label{eq:ccc} \\
\beta &\equiv & \frac{c_2-c_1}{c_2+c1}~,\quad -1\leq \beta \leq 1,
\label{eq:beta}
\end{eqnarray}
and $\mbox{sign}(\omega )=\pm 1$ for $\omega >0$ and $\omega <0$, respectively.
The parameters $c$ and $\beta $ \cite{error} are determined by the
asymptotic behaviour (\ref{eq:asymasym}) of the probability density. The
value of $\beta $ depends on the asymmetry of the probability density $f(x)$.

The final form (\ref{eq:final}) is very convenient. If we consider a random
variable $x_1$ shifted with respect to $x$, $x_1=x-a$ ($a$ is an arbitrary
number here), its characteristic function would differ from that of $x$
by a simple phase factor, $\tilde f_1(\omega )=e^{i\omega a}\tilde f(\omega )$.
On the other hand, the asymptotic behaviour of the probability density,
Eq. (\ref{eq:asymasym}) is not affected by this transformation. Therefore,
the phase factor in Eq. (\ref{eq:final}) can always be eliminated by this
simple shift.

For $0<\alpha <1$ in Eq. (\ref{eq:asymasym}) we re-write the Fourier transform
as
\begin{equation}\label{eq:expand1}
\tilde f(\omega )=1+\int (e^{-i\omega x}-1)f(x)dx~.
\end{equation}
The contribution of positive $x$ to the above integral can be transformed
into
\begin{equation}\label{eq:interm1}
\int _0^{\infty}(e^{-i\omega x} -1)\left[ f(x)-\frac{c_2}
{x^{\alpha +1}}\right]dx +c_2\int _0^{\infty}(e^{-i\omega x} -1)
\frac{dx}{x^{\alpha +1}}.
\end{equation}
Provided the difference between $f(x)$ and $c_2/x^{\alpha +1}$ decreases
as $x^{-\alpha -2}$ or faster, as $x\rightarrow +\infty $, the first integral
can be expanded in powers of $\omega $, with the leading term given by
\begin{equation}
i\omega \int _0^{+\infty}x\left[ f(x)-\frac{c_2}{x^{\alpha +1}}\right]dx~.
\end{equation}
The second integral in (\ref{eq:interm1}) is transformed by variable
substitution $\omega x=it$ (for $\omega >0$) into [cf. Eq. (\ref{eq:Gamint})]
\begin{equation}\label{eq:Gamint1}
c_2e^{i\frac{\pi \alpha }{2}}
\omega ^\alpha \int _0 ^{\infty}t^{-\alpha -1}(e^{-t}-1)dt,
\end{equation}
which again gives the $\Gamma $-function \cite{comment1}. After we apply the
same procedure to the negative-$x$ part of the integral in (\ref{eq:interm1}),
the expansion
of $\tilde f(\omega )$ at small $\omega $ is established in exactly the same
form as that of Eq. (\ref{eq:semif}) (for $\omega >0$). However, for
$0<\alpha <1$ the parameter $a$ is no longer the mean value. Instead,
it is given by
\begin{equation}\label{eq:a}
a=\int _{-\infty }^0 x \left[ f(x)-\frac{c_1}{|x|^{\alpha +1}}\right]dx
+\int _0^{\infty } x \left[ f(x)-\frac{c_2}{x^{\alpha +1}}\right]dx~.
\end{equation}
Also, the next term omitted in Eq. (\ref{eq:semif}) may now be greater than
$O(\omega ^2)$. Nevertheless, the small-$\omega $ behaviour of the
Fourier transform is still represented by Eq. (\ref{eq:final}).

If $\alpha =1$ in Eq. (\ref{eq:asymasym}), the expansion of $\tilde f(\omega )$
also contains $\omega \ln \omega $ terms. In this case it can be presented
as
\begin{equation}\label{eq:alpha2}
\tilde f(\omega )\simeq 1-i\omega a -c|\omega |\left[ 1-i\frac{2}{\pi }
\beta \,\mbox{sign}(\omega ) \ln |\omega |\right]~,
\end{equation}
where $c=\frac{\pi}{2}(c_1+c_2)$, which can be obtained from Eq. (\ref{eq:ccc})
at $\alpha \rightarrow 2$, $\beta $ is given by Eq. (\ref{eq:beta}),
and
\begin{equation}\label{eq:a2}
a=(c_2-c_1)(1-\bbox{C})+\int _{-\infty}^0x \left[ f(x)-\frac{c_1}{1+x^2}\right]
dx +\int _0^{\infty } x \left[ f(x)-\frac{c_2}{1+x^2}\right]dx~,
\end{equation}
where $\bbox{C}\approx 0.577$ is the Euler constant.
Note that if the probability distribution is symmetric asymptotically, i.e.,
$c_1=c_2$, then $\beta =0$, and $a$ in Eqs. (\ref{eq:a}) and (\ref{eq:a2}) is
the mean value calculated in the principal value sense. If the probability
density is fully symmetric, $f(-x)=f(x)$, then $\tilde f(\omega )$ is real,
$a=\beta =0$, and the behaviour of the characteristic function at small
$\omega $ is especially simple:
\begin{equation}\label{eq:fsymm}
\tilde f(\omega )\simeq 1-c|\omega |^\alpha ~.
\end{equation}

After establishing the form of of $\tilde f(\omega )$ at small $\omega $,
Eq. (\ref{eq:final}) for $\alpha \neq 1$, we can proceed to derive the limit
theorem, starting from Eq. (\ref{eq:Fnom}):
\begin{eqnarray}
\tilde F_n(\omega )&=&e^{-i\omega a}\left[ 1-\frac{c \left( 1+i
\beta \,\mbox{sign}(\omega ) \tan \frac{\pi \alpha }{2} \right)n^{1-\alpha }
|\omega | ^\alpha }{n} \right]^n \nonumber \\
&\simeq & e^{-i\omega a}\exp \left[ -c \left( 1+i\beta \,\mbox{sign}(\omega )
\tan \frac{\pi \alpha }{2} \right)n^{1-\alpha }|\omega | ^\alpha \right]~,
\end{eqnarray}
for large $n$ (this formula appears in the theorem by A. Ya. Khintchine
and P. L\'evy as a canonical representation of stable probability
distributions, see Ref. \cite{limit}). Using the last expression in
$F_n(X)=\frac{1}{2\pi} \int e^{i\omega X} F_n(\omega )d\omega $ we obtain
the limit distribution in the following form:
\begin{equation}\label{eq:Fnfab}
F_n(X)=n^{\frac{\alpha -1}{\alpha }}c^{-\frac{1}{\alpha }}f_{\alpha \beta }
\left[n^{\frac{\alpha -1}{\alpha }}c^{-\frac{1}{\alpha }} (X-a)\right] ~,
\end{equation}
where
\begin{equation}\label{eq:fab}
f_{\alpha \beta }(x)=\int _{-\infty }^{\infty }e^{i\omega x-|\omega |^\alpha }
\exp \left[-i\beta \,\mbox{sign}(\omega )\tan \frac{\pi \alpha }{2}
|\omega |^\alpha \right] \frac{d\omega }{2 \pi}~.
\end{equation}
is a universal function of the two parameters, $\alpha $ and $\beta $,
normalized to unity: $\int f_\alpha (x)dx=1$. The results for $\alpha =1$
are obtained in a similar way, with $a$ replaced by $a+c\frac{2}{\pi} \beta
\ln n$ in Eq.~(\ref{eq:Fnfab}), and $f_{1 \beta }(x)$ given by
Eq.~(\ref{eq:fab}), in which $\tan \frac{\pi \alpha }{2}$ is replaced
with $-\frac{2}{\pi}\ln |\omega |$.

Equation (\ref{eq:Fnfab}) shows that for $0<\alpha <1$ the limit distribution
of the average widens with the increase of $n$, i.e., fluctuations of the
average increase with the number of variables averaged. Since
$n^{\frac{\alpha -1}{\alpha }}a\rightarrow 0$ for $n\rightarrow \infty$, the
shift of the distribution (\ref{eq:Fnfab}) by $a$ is actually unimportant
in this case and one can put $a=0$. For $\alpha =1$ the shape of the
distribution $F_n(X)$ does not
depend on $n$, i.e., fluctuations are neither enhanced nor suppressed by
averaging. If $\beta \neq 0$ the whole distribution is gradually shifted
proportionally to $\ln n$ into the direction determined by the sign of $\beta$.
For $1<\alpha <2$ the distribution of the average does become narrower with
$n$, however the rate of suppression of fluctuations, $X\propto
n^{-(\alpha -1)/\alpha }$ is slower than the standard CLT $n^{-1/2}$.
Again, for symmetrically distributed $x_i$, the limit distribution $F_n(X)$
is even simpler, as $a=\beta =0$ in Eqs. (\ref{eq:Fnfab}) and (\ref{eq:fab}).

There are a few cases where $f_{\alpha \beta }$ and, hence, $F_n(X)$, are
known explicitly. For $\alpha =1$, $\beta =0$ ($c_1=c_2$) Eq. (\ref{eq:fab})
gives the Cauchy law,
\begin{equation}\label{eq:Cauchy}
f_{1,0}(x)=\frac{1}{\pi}\,\frac{1}{1+x^2}~,
\end{equation}
and $c=\frac{\pi}{2}(c_1+c2)$. For $\alpha =1/2$, $\beta =0$ the limit
function can be expressed in terms of the error function
$\Phi (s)=2\pi ^{-1/2}\int _0^s e^{-t^2}dt$ \cite{Ryzhik}:
\begin{equation}\label{eq:f1/2}
f_{\frac{1}{2},0}(x)=-\frac{1}{2\sqrt{\pi}|x|^{3/2}}
{\rm Im}\left\{ e^{-i\frac{\pi}{4}}  e^{-\frac{i}{4x}}\left[ 1 - \Phi \left( 
\frac{1}{2\sqrt{ix}}\right) \right] \right\} .
\end{equation}
For the same $\alpha =1/2$ in the maximally asymmetric case, $c_1=0$,
$c_2>0$, i.e., $\beta =1$, which takes place if the random variables $x_i$
are positive, one easily obtains the following simple answer
\cite{limit}:
\begin{eqnarray}\label{eq:smirnov}
f_{\frac{1}{2},1}(x)=\cases {0,\quad x<0, \cr
(2\pi )^{-1/2}e^{-1/2x}x^{-3/2},\quad x>0, \cr }
\end{eqnarray}

%*******************************************************************

%*******************************************************************

\begin{references}
\bibitem[*]{affil}Present address: Department of Applied Mathematics and
Theoretical Physics, The Queen's University of Belfast, Belfast BT7 1NN, UK.

\bibitem{Flambaum94}  V. V. Flambaum and G. F. Gribakin, Phys. Rev. C {\bf 50},
3122 (1994).

\bibitem{Sushkov}  O. P. Sushkov and V. V. Flambaum, Pis'ma Zh. Eksp. Teor.
Fiz. {\bf 32}, 377 (1980) [JETP Letters {\bf 32}, 352 (1980)].

\bibitem{limit} B. V. Gnedenko and A. N. Kolmogorov {\it Limit Distributions
for Sums of Independent Random Variables} (Addison-Wesley, Cambridge, 1954).

\bibitem{levy}  J-P. Bouchard and A. Georges, Phys. Reports, {\bf 195},
127 (1990).

\bibitem{Landau:77}
L. D. Landau and E. M. Lifshitz, {\it Quantum Mechanics}, 3rd ed. (Pergamon
Press, Oxford, UK, 1977).

\bibitem{BM}A. Bohr and B. Mottelson, {\em Nuclear structure, Vol. 1}
(Benjamin, New York, 1969).

\bibitem{Brody}T. A. Brody, J. Flores, J. B. French, P. A. Mello, 
A. Pandey and S. S. M. Wong, Rev. Mod. Phys. {\bf 53}, 385 (1981).

\bibitem{comment}The probability of having two $p$ resonances close together
at some energy $E$ is small because of level repulsion, and if a $s$ resonance
occurs close to $E$, the asymmetry is small. Therefore, such events do not
influence the asymptotic behaviour of the corresponding probability density.

\bibitem{Flambaum83}  V. V. Flambaum and O. P. Sushkov, Nucl. Phys. {\bf A412},
13 (1984).

\bibitem{error} This derivation uncovers an error in Ref. \cite{limit}.
To be consistent with the authors' definition of the asymptotic form of the
distribution (see Theorem on p.~164 and Eq. (24) of \S ~37, p.~182),
the expressions for the parameter $c$ of the characteristic function
of the stable distribution on pp.~169 and 170 must contain an
additional factor $\alpha $.

\bibitem{Ryzhik}I. S. Gradshteyn and I. M. Ryzhik, {\em Table of Integrals,
Series, and Products} (Academic Press, London, 1980).

\bibitem{comment1}The standard integral representation
$\Gamma (z)=\int _0^\infty t^{z-1}e^{-t}dt$ defines the $\Gamma $-function on
the real positive axis, $z>0$. For $z<0$, $\Gamma (z)$ can be defined on
every interval $-(m+1)<z<-m$, where $m$ is a non-negative integer, through the
convergent integral
$\Gamma (z)=\int _0 ^\infty t^{z-1}\left[ e^{-t}-\sum _{k=0}^{m}
(-t)^n/n!\right] dt$.

\end{references}
\end{document}